\documentclass[aps,prd,groupedaddress,nofootinbib,twocolumn]{revtex4}
\usepackage{graphicx,amssymb,amsmath,bm}
\usepackage{dsfont} 
\usepackage{caption}
\usepackage[caption=false]{subfig}
\usepackage{multirow}

\newcommand{\be}{\begin{equation}}
\newcommand{\ee}{\end{equation}}
\newcommand{\bea}{\begin{eqnarray}}
\newcommand{\eea}{\end{eqnarray}}
\newcommand{\nab}{\nabla}
\newcommand{\nn}{\nonumber}
\def\rf#1{(\ref{#1})}

\usepackage{accents}
\newlength{\dhatheight}

\newcommand{\bra}[1]{\left(#1\right)}

\newcommand{\brac}[1]{\left\{#1\right\}} 
\newcommand \veps {\varepsilon}
\newcommand{\A}{{\cal A}}
\newcommand{\E}{{\cal E}}
\renewcommand{\H}{{\cal H}}
\newcommand{\K}{{\cal K}}
\newcommand{\M}{\mathbb{M}}

\begin{document}

\title{The Covariant Tolman-Oppenheimer-Volkoff Equations I: the Isotropic Case}

\author{Sante Carloni}
\author{Daniele Vernieri}
\address{ Centro Multidisciplinar de Astrof\'{\i}sica - CENTRA,
Instituto Superior Tecnico - IST,
Universidade de Lisboa - UL,
Avenida Rovisco Pais 1, 1049-001, Portugal}
\begin{abstract}
We construct a covariant version of the Tolman-Oppenheimer-Volkoff equations in the case of isotropic sources. The new equations make evident the mathematical problems in the determination of interior solutions of relativistic stellar objects. Using a reconstruction algorithm we find two physically interesting generalisations of previously known stellar interior solutions. The variables that we use also allow an easier formulation of known generating theorems for solutions associated to relativistic stellar objects.
\end{abstract}
\maketitle

\tolerance=5000
\section{Introduction} 
Covariance is one of the most fundamental and important aspects of General Relativity (GR). It allows to explore many crucial properties of curved spacetime writing equations that are independent of the choice of a specific observers. Nevertheless, much of the research in relativistic gravitation is performed giving up covariance and using only  a given set of coordinates. While this is perfectly legitimate, it also limits our understanding of a given gravitational system. A classical example is the case of the actual nature of the singularity in the Schwarzschild solution, which was understood to be an artefact of the coordinate system only well after its initial derivation. Issues like these motivate the development of formalisms that preserve as much as possible the covariant character of the Einstein field equations. 

One of these methods, nowadays dubbed ``$1+3$ covariant approach'', has been proposed by Ehlers, Ellis and others to treat cosmological spacetime~\cite{Covariant}. The $1+3$ covariant approach has been very successful not only in clarifying  aspects of the usual Friedmann-Robertson-Walker spacetimes but also in the case of the more complicated Bianchi models and even inhomogeneous cosmologies.

The $1+3$ approach makes full use of the symmetries of the spacetime in which it is employed, and more specifically of the concept of  foliation of these spacetimes. This, however, is also a limitation. Applying the $1+3$ approach to less symmetric spacetimes considerably reduces the advantages of this formalism.  Recently, a new formalism adapted to Locally Rotational Symmetric (LRS) spacetimes was proposed. Since these spacetimes are locally symmetric around a specific spacelike direction, the new approach was called ``$1+1+2$ covariant approach''~\cite{Clarkson:2002jz,Betschart:2004uu,Clarkson:2007yp}. 

The $1+1+2$ covariant approach can be applied to any LRS spacetimes and in particular to a subclass of these spacetimes (LRSII) which contains static spherically symmetric metrics. In astrophysics these metrics are relevant for black holes in the vacuum case, but also, in the non vacuum case, to describe the interior solutions of relativistic stars.

This last class of systems is commonly investigated by rearranging of the Einstein field equations and the Bianchi Identities into the well known Tolman-Oppenheimer-Volkoff (TOV) equations. The TOV equations  characterise the pressure profile of a static and spherically symmetric object made of a given matter fluid. Since the proposal of the TOV equations (but also before that~\cite{Tolman:1939jz}), much work has been done to look for exact and numerical solutions to model the interior of  relativistic stars. In addition, in recent years some general properties of the exact solutions of these equations have been discovered (see {\it e.g.} Refs.~\cite{Boonserm:2005ni,Boonserm:2006up,ExactSolutions,Baumgarte:1993,Mak:2013pga}). However, in spite of the relevance of these results, the actual resolution of the problem of stellar interior is still a formidable task and it is largely hindered by a number of technical difficulties. 

The aim of this paper is to apply the $1+1+2$ covariant approach to the problem of the determination of the interior solutions of isotropic relativistic stars. We will construct a covariant version of the TOV equations and we will use the advantages of the covariant approaches to have a more clear understanding of the mathematical structure of these equations, the most convenient resolution strategies and some peculiarities of the solutions obtained. The character of the present work will be mainly foundational. The new TOV equations will have straightforward generalisation to a series of non trivial cases, which will be considered in following works.

The paper is organised as follows. In Section II we will introduce the basics of $1+1+2$ formalism. In Section III we will derive the key $1+1+2$ equations and we will construct a set of variables which are more suited to work with the TOV equations. We will comment also on the junction conditions in these variables. In Section IV we will give the covariant TOV equations and suggest some resolution methods. In Section V we will present a reconstruction algorithm to generate some new solutions. In Section VI we will use the covariant TOV equations to derive some known generating theorems \cite{Boonserm:2005ni} and to prove new ones. Section VII is dedicated to the conclusions.

Unless otherwise specified, natural units ($\hbar=c=k_{B}=8\pi G=1$) will be used throughout this paper and Latin indices run from 0 to 3. The symbol $\nabla$ represents the usual covariant derivative and $\partial$ corresponds to partial differentiation. We use the $-,+,+,+$ signature and the Riemann tensor is defined by
\begin{equation}
R^{a}{}_{bcd}=\Gamma^a{}_{bd,c}-\Gamma^a{}_{bc,d}+ \Gamma^e{}_{bd}\Gamma^a{}_{ce}-\Gamma^e{}_{bc}\Gamma^a{}_{de}\;,
\end{equation}
where the $\Gamma^a{}_{bd}$ are the Christoffel symbols (i.e. symmetric in the lower indices), defined by
\begin{equation}
\Gamma^a_{bd}=\frac{1}{2}g^{ae}
\left(g_{be,d}+g_{ed,b}-g_{bd,e}\right)\;.
\end{equation}
The Ricci tensor is obtained by contracting the {\em first} and the {\em third} indices
\begin{equation}\label{Ricci}
R_{ab}=g^{cd}R_{acbd}\;.
\end{equation}
Symmetrisation and the anti-symmetrisation over the indexes of a tensor are defined as 
\begin{equation}
T_{(a b)}= \frac{1}{2}\left(T_{a b}+T_{b a}\right)\;,~~ T_{[a b]}= \frac{1}{2}\left(T_{a b}-T_{b a}\right)\,.
\end{equation}
Finally the Hilbert--Einstein action in the presence of matter is given by
\begin{equation}
{\cal A}=\frac12\int d^4x \sqrt{-g}\left[R+ 2{\cal L}_m \right]\;.
\end{equation}

\section{$1+1+2$ Covariant Approach} 

Our discussion will be based on the $1+1+2$ Covariant Approach~\cite{Clarkson:2002jz,Betschart:2004uu,Clarkson:2007yp}. This semi-tetradic formalism is built on a  threading decomposition of the spacetime manifold with respect to a timelike congruence  and a chosen spacelike vector field orthogonal to this congruence. As a consequence, one can construct a set of tensorial objects (the {\it $1+1+2$ variables}) with a rigorous mathematical definition and a clear physical meaning. The Bianchi and Ricci identities can be then written as a closed system of propagation and constraint equations which  are completely equivalent to the Einstein field equations. Here we give a short summary of the main aspects of the $1+1+2$ formalism. A more complete description can be found in Refs.~\cite{Clarkson:2002jz,Betschart:2004uu,Clarkson:2007yp}.

The construction of the $1+1+2$ variables and their equations  starts with  the definition of a vector field $u^a$ associated to a congruence  time-like  integral curves  $(u_{a} u^{a} = -1)$  and a vector $e_a$ associated to a congruence of spacelike $(e_{a} e^{a} = 1)$  integral curves. The geometry of the orthogonal hypersurfaces is defined by the two projection tensors
\begin{equation}\label{projT}
\begin{split}
&h^a{}_b=g^a{}_b+u^au_b~,~~h^{a}{}_{a} = 3~, \\
&N_{a}{}^{b} \equiv h_{a}{}^{b} - e_{a}e^{b} = g_{a}{}^{b} + u_{a}u^{b} 
- e_{a}e^{b}~,~~N^{a}{}_{a} = 2~, 
\end{split}
\end{equation}
which represent the metric of the 3-spaces  orthogonal to $u_a$ and of 2-spaces ($W$) orthogonal to $u_a$ and $e_a$ respectively.  Each of these hypersurfaces posses a volume form given by
\begin{equation}
\begin{split}
&\veps_{abc}= \eta_{dabc}u^{d}~, \\
&\veps_{ab}\equiv\veps_{abc}e^{c}~, 
\end{split}
\end{equation}
which can be used to isolate vortical contributions.

Using $u_a$, $e_a$, $h_{ab}$,   $N_{ab}$ any tensorial object can be split according to the above foliations. For example, any 4-vector $X^{a}$ can be irreducibly split as 
\bea
&&X^{a} =  \Xi_0 u^{a}+ \Xi_1 e^{a} + \Xi^{a}_2\,, \\ 
&&\Xi _0=X^{a}u_a \,, \quad \Xi_1 \equiv X_{a}e^{a} \,, \quad\Xi ^{a} \equiv N^{ab}X_{b}~. 
\label{equation1} 
\eea
In the case of  a symmetric 4-tensor, $X^{ab}$ the decomposition is longer:
\bea 
&&\nn X^{ab} = \Xi _0 u^a u^b +\Xi_1 e^{a}e^{b}+\Xi_2 N^{ab}+2 \Xi_1^{(a}u^{b)}\\ 
&&~~~~~~~~+2 \Xi_2^{(a}e^{b)} +2 \Xi_2 u^{(a}e^{b)}+\Xi^{ab}~,
\eea 
where 
\bea
\begin{split}
&\Xi _0 =X_{ab}u^{a}u^{b}, \\
&\Xi_1 \equiv X_{ab}{e}^{a}{e}^{b}~,\\
&\Xi_2 \equiv  \frac{1}{2}X_{ab}N^{ab}~,\\
&\Xi_1^{a} \equiv X_{cd}N^{a}{}_{c}u^{d}~,\\
&\Xi_2^{a} \equiv X_{cd}N^{a}{}_{c}e^{d}~,
\nonumber \\
&\Xi^{ab}\equiv \mathds{X}^{\{ab\}}~,\nonumber
\end{split}
\eea
and the curly brackets denote the projected symmetric trace free part of a tensor with respect to $e^{a}$: 
\be
 \mathds{X}^{\{ab\}} 
\equiv \bra{ N^{c}{}_{(a}N_{b)}{}^{d} - \frac{1}{2}N_{ab} N^{cd}} \mathds{X}_{cd}~.
\ee

The same type of decomposition can be applied to the covariant derivative vector to obtain a set of different derivative operators: the \textit{covariant time derivative} 
\be
\dot{X}^{a..b}{}_{c..d}{} = u^{e} \nab_{e} {X}^{a..b}{}_{c..d}~, 
\ee
the full \textit{orthogonally projected covariant derivative} $D$, 
\be
D_{e}X^{a..b}{}_{c..d}{} = h^a{}_f h^p{}_c...h^b{}_g h^q{}_d h^r{}_e \nab_{r} {X}^{f..g}{}_{p..q}~,
\ee
the \textit{hat-derivative} 
\be
\hat{X}_{a..b}{}^{c..d} \equiv  e^{f}D_{f}X_{a..b}{}^{c..d}~, 
\ee
i.e.  the component of $D$  along the $e^a$ vector-field, and the \textit{$\delta$ -derivative}
\be
\delta_fX_{a..b}{}^{c..d} \equiv  N_{a}{}^{f}...N_{b}{}^gN_{h}{}^{c}..
N_{i}{}^{d}N_f{}^jD_jX_{f..g}{}^{i..j}\;,
\ee
which is the projected derivative onto $W$.

The $1+1+2$ variables can be defined using the decomposition of the covariant derivative of $u_a$ and the orthogonally projected covariant derivative of the vector field $e_a$:
\begin{equation}\label{112Var}
\begin{split}
\nabla_{a}u_{b} =& u_{a}(\A e_{b}+ \A_{b})+ \frac{1}{3}\left(\tilde{\theta}+\bar{\theta}_u\right)  \left(N_{ab}+e_a e_b\right) \\
&+\Sigma\bra{ e_ae_b - \frac{1}{2}N_{ab}} + 
2\Sigma_{(a}e_{b)} + \Sigma_{ab} \\
&+ 2\veps_{cab} (\Omega e^{c} +\Omega^{c}) ~, \\
{\rm D}_{a}e_{b} =& e_{a}a_{b} + \frac{1}{2}\phi N_{ab}+ \zeta_{ab} + 
\xi\veps_{ab} ~, 
\end{split}
\end{equation}
where
\be
\begin{split}
& \A=e_a \dot{u}^{a}~, \quad  \A^a=N_{ab}\dot{e}^{b}~,\\
& \tilde{\theta}=\delta_a u^a~,\quad \bar{\theta}=a_{b}u^b,\quad\tilde{\theta}+\bar{\theta}=D_a u^a=\Theta~,\\
&\Omega =\frac{1}{2}\veps^{abc}D_{[a} u_{b]} e_a~, \quad\Omega^{a}=\frac{1}{2}\veps^{abd}D_{[a} u_{b]} N_d{}^{a}~,\\
&\Sigma=\sigma^{ab}\bra{ e_ae_b - \frac{1}{2}N_{ab}}~,\quad \Sigma_a=\sigma_{cd} e^c N^{d}{}_{a}~,\\
&\Sigma_{ab}= \mathds{\sigma}_{\{cd\}}~,\\
&\sigma_{ab}= \left(h^{c}{}_{(a}h_{b)}{}^{d} - \frac{1}{3}h_{ab} h^{cd}\right)D_{c}u_{d}~,\\
&a_{b} =  e^{c}{\rm D}_{c}e_{b} = \hat{e}_{b}~, \qquad \phi = \delta_a e^a~, \\  
&\zeta_{ab} =  \delta_{ \{c}e_{d\}}~, \quad
\xi =  \frac{1}{2} \veps^{ab}\delta_{a}e_{b}~.
\end{split}
\ee
The set is completed by  some additional variables related to the decomposition of electric and magnetic parts of the Weyl curvature tensor $C_{abcd}$:
\be
\begin{split}
&\E=C^{ab}{}_{cd}u^{c}u^{d}\bra{ e_ae_b - \frac{1}{2}N_{ab}}~,\\
&\E_a=C_{cd ef}u^{e}u^{f} e^c N^{d}{}_{a}~,\quad \E_{ab}= C_{\{ab\}cd}u^{c}u^{d},\\
&\H=\frac{1}{2}\veps^{a}{}_{de}C^{de b}{}_{c}u^c\bra{ e_ae_b - \frac{1}{2}N_{ab}}~,\\
&\H_a=\frac{1}{2}\veps_{cfe}C^{fe}{}_{dh}u^h e^c N^{d}{}_{a}~,\\
&\H_{ab}=\frac{1}{2}\veps_{\{ade}C^{de}{}_{b\}c}u^c~.
\end{split}
\ee
For an observer that moves on the geodesic congruence defined by $u_a$, the expansion  of the geodesics will be  given by $\Theta=\tilde{\theta}+\bar{\theta}~$, the deviation from free fall will be represented by the  components $\A$  and $\A^{a}$ of  the acceleration vector $\dot{u}_a$,  the components $\Sigma_{ab}$, $\Sigma_{a}$ and $\Sigma$ of the  shear $\sigma_{ab}$ will represent the non isotropic deformation of the geodesic flow and the components  $\Omega$ and $\Omega_a$ of the vorticity $\omega_{ab}=D_{[a} u_{b]}$ its rotation. Similarly if the same observer chooses $ e^{a} $ as special direction in the spacetime, $\phi$ represents the expansion of the integral curves of the vector field $ e^{a} $,  $\zeta_{ab}$ is their distortion  ({\it i.e.} the \textit{shear of $e^{a}$}) and $a^{a}$ the change of the vector $e_a$ along its integral curves ({\it e.g.} its \textit{acceleration}). We can also interpret $\xi$ as a representation of the ``twisting'' or rotation of the integral curves of $e_a$ ({\it i.e.} the \textit{vorticity} associated with $e^{a}$). Finally $\E, \E_a, \E_{ab}$ are related wth the Newtonian part of the gravitational potential while $\H, \H_a, \H_{ab}$ are related to tidal relativistic forces.

The matter stress energy tensor $T^{m}_{ab}$ can also be decomposed with respect to $u_a$, $e_a$, $N_{ab}$ to give
\bea 
&&\nn T_{ab} = \mu  u_{a}u_{b} + \left(p+\Pi\right)e_{a}e_{b}+\left(p-\frac{1}{2}\Pi\right)N_{ab}\\ 
&&~~~~~~~+2 Q e_{(a}u_{b)}+ 2Q_{(a}u_{b)}+2\Pi_{(a}e_{b)}+\Pi_{ab}\,,\label{Tab112}
\eea 
which defines the matter $1+1+2$ variables as
\be
\begin{split}
& \mu=T_{ab}{u}^{a}{u}^{b}~,\\
&p=\frac{1}{3}T_{ab}\left({e}^{a}{e}^{b}+N^{ab}\right)~,\\
& \Pi=\frac{1}{3}T_{ab}\left(2{e}^{a}{e}^{b}-N^{ab}\right)~,\\
&Q=\frac{1}{2}T_{ab}e^{a}u^{b}~,\\
&Q_a=T_{cd}N^{a}{}_{c}u^{d}~,\\
&\Pi_{a}=T_{cd}N^{a}{}_{c}e^{d}~,\\
&\Pi_{ab}=T_{\{ab\}}~,\\
& p_r=p+\Pi=T_{ab}{e}^{a}{e}^{b}~,\\
& p_\perp=p-\frac{1}{2}\Pi=\frac{1}{2}T_{ab}N^{ab}~.
\end{split}
\ee
The last two equation connect the 1+1+2 variables to  the radial and transversal pressure $p_r$ and  $p_\perp$, {\it i.e.} the components of the fluxes that would appear in the left hand side of the radial and angular Einstein equations. 

\section{$1+1+2$ equations for static and spherically symmetric spacetimes} 

The 1+1+2 formalism  is able to describe in a natural way all LRS spacetimes in which one can define covariantly a unique, preferred spatial direction. In the following we are interested in the case of the rotation free, static and spherically symmetric spacetimes (LSRII). In this case \textit{all} the $1+1+2$ vectors and tensors vanish, as well as the variables $\Omega$, $ \xi $, $ \H $, $\Theta$, $\Sigma$ and $Q$. Thus one is left with the six scalars $\brac{\A, \phi, \E, \mu, p, \Pi }$. In this work we will also assume that the source for the gravitational field is a completely isotropic fluid, {\it i.e.} $\Pi =0$.

The set of $1+1+2$ equations which describe spherically symmetric static spacetimes is~\cite{Clarkson:2002jz,Betschart:2004uu,Clarkson:2007yp,Nzioki:2009av}:
\bea 
\hat\phi &= -&\frac12\phi^2 -\frac23\mu-\E~,
\label{StSpSymEqGen1}\\
\hat\E -\frac13\hat\mu &=-& \frac32\phi \E~,
\label{StSpSymEqGen2}\\
\hat p&= &-\bra{\mu+p}\A~,
\label{StSpSymEqGen3}\\
\hat\A &= -&\bra{\A+\phi}\A + \frac12\bra{\mu +3p}~. 
\label{StSpSymEqGen4}
\eea
with the constraint
\begin{equation}
0 = - \A\phi + \frac13 \bra{\mu+3p} -\E ~.
\label{ConstSpSymEqGen}
\end{equation}
In order to solve the equations above it is useful to define the Gaussian curvature $K$ of $W$~\cite{Clarkson:2002jz,Betschart:2004uu,Clarkson:2007yp}
\be
K = \frac13 \mu - \E + \frac14 \phi^{2}  
\label{GaussCurv}.
\ee
The  propagation equation for $K$ can be then written as 
\bea
\hat{K} = -\phi K. \label{propGauss}
\eea
This last equation is the starting point for the choice of an affine parameter related to the {\it hat} derivative which can lead to a  simplification of the final  $1+1+2$ equations. For our purposes a convenient parameter to choose the logarithmic space variable $\rho$ such that  $\hat{K} = K_{,\rho}\phi$~\cite{Carloni:2013ite}. This operation allows to make the {\it hat} derivatives dimensionless. In this way Eq.~\rf{propGauss} becomes  
\begin{equation}
K_{,\rho}=-K \label{eqKRho}~,
\end{equation}
and the other equations become 
\bea 
&&
\phi\,\phi_{,\rho}= -\frac12\phi^2 -\frac23\mu-\E~,
\label{StSpSymEqGen1M}\\
&&
\E_{,\rho}-\frac13\mu_{,\rho} =-\frac{3}{2} \E~,
\label{StSpSymEqGen2M}\\
&&
\phi\,p_{,\rho} = -\bra{\mu+p}\A~,
\label{StSpSymEqGen3M}\\
&&
\phi\, \A_{,\rho}= -\bra{\A+\phi}\A + \frac12\bra{\mu +3p}~,
\label{StSpSymEqGen4M}\\
&&  \A\,\phi - \frac13 \bra{\mu+3p} +\E ~=0~,\label{ConstSpSymEqGenM}\\
&& K = \frac13 \mu - \E + \frac14 \phi^{2} \label{GaussCurvRho}.
\eea
Notice that in the system above the Eq.~\rf{eqKRho} is decoupled. Therefore we cannot eliminate Eq.~\rf{GaussCurvRho} as there can be solutions which satisfy all the  above differential equations, but not Eq.~\rf{GaussCurvRho}.

Another choice can be to use the so-called {\it area radius} for which $K\propto r^{-2}$. It is not difficult to show that  
\be\label{RhoR}
\rho=2\ln\left(\frac{r}{r_0}\right),
\ee 
where $r_0$ is an arbitrary constant. In the rest of this work we will perform the calculations in $\rho$, but we will give the final results in terms of $r$ in such a way to facilitate the connection with known results.

Eqs.~(\ref{StSpSymEqGen1M}-\ref{GaussCurvRho}) characterize completely the static and spherically symmetric metrics in GR, and we can use them to find solutions of Einstein theory with this symmetry. However, as shown in Ref.~\cite{Carloni:2014rba}, this system of equations can be further simplified to a set of dimensionless equations written in terms of variables that help highlighting the physical aspects of the solutions. These variables are defiened as
\begin{eqnarray}
&&\nn X=\frac{\phi_{,\rho}}{\phi}~,\qquad Y= \frac{\A}{\phi}~, \qquad \K=\frac{K}{\phi^2}~,  \\
&&E=\frac{\E}{\phi^2}~, \qquad \mathbb{M}=\frac{\mu}{\phi^2}~, \qquad P=\frac{p}{\phi^2}~.\label{Var}
\end{eqnarray}
Using the affine parameter $\rho$, the $1+1+2$ equation take the form~\cite{Carloni:2014rba}
\bea
&& Y_{,\rho}=\mathbb{M}+3 P-2Y (X+ Y+1)\,,  \label{NewEqGenV1}\\
&& \K_{,\rho}=-\K(1+2X)\,, \label{NewEqGenV2}\\
&&  P_{,\rho}=-2 Y \mathbb{M}-2 P(2X+Y)\,, \label{NewEqGenV3}
\eea
with the constraints
\begin{eqnarray}
&& 2 \mathbb{M}+2 P+2X-2 Y+1=0\,,\label{NewConstrGenV1}\\
&&1- 4\K-4P+4Y=0\,, \label{NewConstrGenV2}\\
&&2 \mathbb{M}+6 P-6 Y-6 E=0\,. \label{NewConstrGenV3}
 \end{eqnarray}
We will make use of the above equations to derive a covariant version of the TOV equations.
 
\subsection{Relation with the metric coefficients} 
As already said, the geometrical variables given above are tensors for the group of transformations that preserve the foliations. At any moment one can break covariance and write the equivalent of these variables in a given system of coordinates in terms of metric coefficients and their derivatives~\cite{Nzioki:2009av}. Such operation shows that the above construction can be viewed as a method to find an `optimal' combination of these quantities which, as we will show in the following, simplifies considerably a number of issues associated to spherically symmetric spacetimes.

Let us consider a generic choice of spherical coordinates. In this case the line element is 
\be\label{sphsymmmetric}
ds^{2}= - A(p)dt^{2} + B(p)dp^{2} + C(p)(d\theta^{2} + \sin^2\theta d\phi^{2})\,,
\ee 
and we have: 
\be
\begin{split}
&u_a= \left(\sqrt{A(p)},0,0,0\right)\,,\\
& e_a= \left(0,\sqrt{B(p)},0,0\right)\,,\\ 
&N_{a}{}^{b}= C(p)\left(\delta_{a \theta}\delta^{b}{}_{ \theta}+\delta_{a \phi}\delta^{b}{}_{\phi}\right)\,.
\end{split}
\ee
This implies that $\A$, $\phi$ and $K$ are given by 
\be
\A=\frac{1}{2A\sqrt{B} }\frac{d A}{d p}\,, 
\label{Apo}
\ee
\be
\phi=\frac{1}{C\sqrt{B} }\frac{d C}{dp}\,,
\label{Phip}
\ee
\be
K=\frac{1}{C}\,,
\label{Kp}
\ee
while
\be
Y=\frac{1}{2}\frac{C}{A}\frac{A_{,p}}{C_{,p}}\,, 
\label{Yp}
\ee
\be
\K=\frac{C B}{C_{,p}^2}\,.
\label{Kap}
\ee
It is not too difficult to give the equivalent of these expressions in terms of the affine parameter $\rho$ or $r$. In terms of  $\rho$ and using the same procedure one finds that 
\be
\A=\frac{1}{2A\sqrt{B} }\frac{d A}{d \rho}\,, 
\label{AsolutionRho}
\ee
\be
\phi=\frac{1}{\sqrt{B}}\,,
\label{PhisolutionRho}
\ee
\be
K=\frac{1}{C_0} e^{-\rho}=K_0 e^{-\rho}\,,
\label{KsolutionRho}
\ee
\be
Y=\frac{1}{2}\frac{A_{,\rho}}{A}\,, 
\label{YRho}
\ee
\be
\K=K_0 B(\rho) e^{-\rho}\,.
\label{KaRho}
\ee
Notice that dimensional consistency requires  the the constant $K_0$ must have dimensions of the inverse of a length square.
In terms of $r$ the $1+1+2$ variables read  
\be
\A=\frac{1}{2 A \sqrt{B} }\frac{d A}{d r}~,
\label{Asolutionr}
\ee
\be
\phi=\frac{2}{r\sqrt{B}}~,
\label{Phisolutionr}
\ee
\be
K=\frac{\bar{K}_0}{r^{2}}\,,
\label{Ksolutionr}
\ee
\be
Y=\frac{1}{4}\frac{A_{,r}r}{A}\,, 
\label{Yr}
\ee
\be
\K=\bar{K}_0 B(r)\,.
\label{Kar}
\ee
The constant $\bar{K_0}$ is dimensionless and can assume in principle any numerical value, but in order to obtain the true area radius one has to set $\bar{K}_0=1$.  Comparing the  expressions for  $K$ in $\rho$ and $r$ and using eq. \eqref{RhoR}, one obtain that $K_0=\bar{K}_0 r_0^{-2}= r_0^{-2}$.  In the following we will use these relations to present solution in either of the $\rho$ and $r$ coordinate systems.

The expressions above, combined with the constraint in Eq.~\rf{NewConstrGenV2}, show that there exists a relation between the zeros and the singular points of the metric coefficients and the zeros and divergences of the pressure. For example, a zero in  $A$ with a regular $B$  leads to  divergence in $Y$ and therefore in $P$ and $p$. This feature will be useful when we will define a reconstruction algorithm.

The above results can be used at any stage of the calculations to give the results in coordinates. We will also employ them  in the next subsection, to give the $1+1+2$ formulation of the junction conditions.
\subsection{Junction conditions} 
In the following we will calculate a number of static spherically symmetric solutions of the Einstein equation with isotropic  sources. Since we will consider these solutions as interior configurations of stellar objects, an important role in the determination of their physical character is played by the way in which they can be matched with a Schwarzschild exterior. In order to achieve this task we will make use of the $1+1+2$ version of the junction conditions given by Israel in Refs.~\cite{Israel:1966rt,Barrabes:1991ng}. 

In the present case, the surface of separation will be the 3-surfaces $\mathcal{S}$ normal to $e_a$. Indicating the jump along $\mathcal{S}$ as $[X]=X^+ -X^-$, the junction conditions read 
\begin{equation}
[g^\mathcal{S}_{ab}]=[N_{ab}+u_{a}u_{b}]= 0\,, \qquad [\bar{K}_{cd}]=0\,,
\end{equation}
where $g^\mathcal{S}_{ab}$ is the metric of $\mathcal{S}$ and $\bar{K}_{cd}$ is the extrinsic curvature of $\mathcal{S}$:
\begin{equation}
\begin{split}
\bar{K}_{ab}=&(N_{a}{}^{c}+u_{a}u^{c})(N_{b}{}^{d}+u_{b}u^{d} )\nabla_{c}e_{d}~.
\end{split}
\end{equation}
 which in the spherically symmetric case  is given by:
 \begin{equation}
\bar{K}_{ab}=\frac{1}{2}\phi N_{ab}+u_au_b \A~,
\end{equation}
the conditions above imply the following junction conditions
\begin{equation}
[u_a]=0\,, \qquad [N_{ab}]=0\,,
\end{equation}
and
\begin{equation}\label{JC2}
[\A]=0\,, \qquad [\phi]=0\,.
\end{equation}
As from the original Israel conditions, if Eqs. \eqref{JC2} are not satisfied, a thin shell  with stress energy tensor
\be 
T^\mathcal{S}_{ab}=(N_{ab}+u_{a}u_b)[\bar{K}]-[\bar{K}_{cd}],
\ee
will be present in the spacetime. Let us look at the Israel's Junction conditions in terms of the variables in Eq.~\rf{Var}. Since $[\phi]=0$ then
\begin{equation}
\left[\frac{ K_0 e^{-\rho}} {\phi^2}\right]=[\K]=0,
\end{equation}
and using the constraint in Eq.~\rf{NewConstrGenV2} above
\begin{equation}\label{PreJunction112}
0=[\K]= \left[P-Y\right].
\end{equation}
Now, since $[\A]=0$, we have $[Y]=0$ and the \eqref{PreJunction112} reduces to 
\begin{equation}\label{JunCov}
 \left[P\right]=0.
\end{equation}
As $[\phi]=0$, this is equivalent to say that $[p]=0$. In other words, in order to provide a smooth junction with the Schwarzschild exterior metric one has to seek a value of the radius in which the pressure is zero.   

It is instructive to break covariance to see how the conditions on the metric are expressed in coordinates.  Eqs.~(\ref{AsolutionRho}-\ref{KaRho}) and~(\ref{Asolutionr}-\ref{Kar}) show that the junction conditions \eqref{PreJunction112} require the continuity of $A$, of its first derivative and also  the continuity of $B$. 
Looking at the radial component of the Einstein equation it becomes clear that this choice sets the radial pressure to zero. Notice that since there is no constraint on the first derivative of $B$  the energy density does not need to be zero. In other words the above conditions imply  
\begin{align}
& \mu\neq 0~,\quad p=0~,
\end{align}
which are the same conclusions of Eq.~\rf{JunCov}.  In the following to ensure the compatibility with a Schwarzschild exterior,  we will  impose directly that $p$ is zero on the boundary of the star. 
\section{$1+1+2$ TOV equations}\label{112TOV}

Considering Eq.~\rf{NewEqGenV2} and eliminating $X$ and $Y$ from Eq.~\rf{NewEqGenV3}, one obtains:
\begin{equation}\label{112TOVPF}
\begin{split}
&P_{,\rho }=-P^2+P \left[
   \mathbb{M}+1-3\left( \mathcal{K}-\frac{1}{4}\right)\right]\\
&~~~~~~~~-\left(\mathcal{K}-\frac{1}{4}\right) \mathbb{M}\,,\\
&\mathcal{K}_{,\rho }=-2 \mathcal{K}\left(\mathcal{K}-\frac{1}{4}-\mathbb{M}\right),
\end{split}
\end{equation}
which are the covariant version of the TOV equations. The system is closed by the equation of state $P=f(\mathbb{M})$. 

If one chooses, as customary, a density profile, the above equations constitutes a set of coupled Riccati and Bernoulli equations.  Although we can solve formally a Bernoulli equation, the same cannot be done for the Riccati one. This suggests that there is no formal general exact solution of the Eqs.~\rf{112TOVPF}. However, one can find a number of  exact solutions in particular cases  (see also Ref.~\cite{Mak:2013pga}). It is important to stress that  these solutions will correspond to actual astrophysical objects only if the following conditions are satisfied (see {\it e.g.} Ref.~\cite{Delgaty:1998uy}): 
\begin{enumerate}
\item $\mu$ and $p$ should be positive inside the object;
\item the gradients of $\mu$ and $p$ should be negative;
\item the speed of sound should be less than the speed of light $0<\frac{\partial p}{\partial \mu}<1$; 
\item the energy conditions should be satisfied;
\end{enumerate}
Remarkably, very few of all the solutions known actually satisfy the above conditions~\cite{Delgaty:1998uy}. In the following we will find also some solutions which are not included in Ref.~\cite{Delgaty:1998uy} and which posses at least a set of their parameters able to satisfy all of these conditions. 

\subsection{Some exact solutions for the perfect fluid case.}\label{SecNewVar}

We will now present briefly some solving strategies for Eqs.~\rf{112TOVPF}. The typical approach is to use and ansatz for the energy density. We will consider here the classic case of a star made of a fluid with constant energy density, which will also allow to  check that the equations are actually correct.

If $\mu$ is constant, $\mathbb{M}$ reads
\begin{equation}
\mathbb{M}= \frac{\mu _0 e^{\rho }}{K_0} \mathcal{K}\,.
\end{equation}
Substituting this into the second of Eqs.~\rf{112TOVPF} this gives the solution for $\mathcal{K}$:
\begin{equation} \label{KMuConstFl}
\mathcal{K}= \frac{3}{3\mathcal{K}_0
   e^{-\rho /2}-\left(3+\mu _0 e^{\rho } r_0^2\right)}~.
\end{equation}
Setting $\mathcal{K}_0=0$ one gets the solution for $P$:
\begin{equation}
P=\frac{\mu _0 e^{\rho } r_0^2
   \left(P_0+3 \sqrt{3-\mu _0 e^{\rho } r_0^2}\right)}{4 \left(\mu _0 e^{\rho }
   r_0^2-3\right) \left(P_0+\sqrt{3-\mu _0 e^{\rho } r_0^2}\right)}~,
\end{equation}
 which, in turn, implies
 \begin{equation}
Y=\frac{\mu _0 e^{\rho } r_0^2}{2 \sqrt{\mu _0 e^{\rho } r_0^2-3}
   \left(P_0+\sqrt{3-\mu _0 e^{\rho } r_0^2}\right)}~.
\end{equation}
The corresponding metric can be written, in terms of the area radious
\begin{subequations}
\begin{align}\label{MuConstFl}
& d s^2 = -Ad t^2+ B d r^2 +r^2\,\big(d\theta^2+\sin^2\!\theta d\phi^2\big)\,,\\
& A =A_0\left(\sqrt{3 - \mu _0 r^2}+P_0\right)^2\,, \label{MuConstFlA}\\
& B =\frac{3}{3 -\mu _0 r^2}\,.
\end{align}
\end{subequations}
The pressure is
\begin{equation}
p=-\frac{\mu _0\left(P_0+3 \sqrt{3-\mu _0 r^2}\right)}{3 \left(P_0+\sqrt{3-\mu _0
   r^2}\right)}\,,
\end{equation}
which, for $P_0<0$, corresponds to the well known result and confirms the correctness of Eqs.~\rf{112TOVPF}.


There are a number of ways in which Eqs.~\rf{112TOVPF} can be solved. An interesting strategy is to choose an ansatz for $\mathcal{K}$ and an equation of state. By setting
\begin{equation}
\begin{split}
&\mathcal{K}= f(P)\,,\\
&\mathbb{M}= g(P)\,,
\end{split}
\end{equation}
and substituting the above relations in the second of Eqs.~\rf{112TOVPF},we obtain the equation
\begin{equation}\label{EqPKM}
\begin{split}
&g=\frac{2  f_{,P}}{4 f}P_{,\rho}+ f-\frac{1}{4},\\
&P_{,\rho } \left[\frac{(4 f-4 P-1) f_{,P}}{8 f}+1\right]+f^2+\left(2
   P-\frac{1}{2}\right) f\\
&~~~~~~+P^2-\frac{3 P}{2}+\frac{1}{16}=0\,.
 \end{split}
\end{equation}
We can choose different forms of the function $f$ such that this equation can be solved exactly.
Let us suppose, for example,
\begin{equation}
f=\frac{1}{4} \left(-4 P\pm4 \sqrt{P}+1\right),
\end{equation}
then Eq.~\rf{EqPKM} gives $P=P_0$  and the function $g$ is 
\begin{equation}
g=-\sqrt{P}-P~.
\end{equation}
Solving for $Y$ one has
\begin{equation}
Y=-\sqrt{P_0}~,
\end{equation}
which correponds to the metric
\begin{subequations}
\begin{align}
& d s^2 = -Ad t^2+ B d r^2 +r^2\,\big(d\theta^2+\sin^2\!\theta d\phi^2\big),\\
& A =A_0 \left(\frac{r_0}{r}\right)^{4 \sqrt{P_0}},\\
& B =\left[1-4
   \left(P_0+\sqrt{P_0}\right)\right],
\end{align}
\end{subequations}
together with the  energy density and pressure given by
\begin{subequations}
\begin{eqnarray}
&& \mu= \frac{ 4\left(\sqrt{P_0}+1\right)
   \sqrt{P_0} }{\left(4 P_0+4 \sqrt{P_0}-1\right) r^2},\\
&&p=-\frac{4 P_0 }{\left(4 P_0+4 \sqrt{P_0}-1\right) r^2},
\end{eqnarray}
\end{subequations}
 and the equation of state
 \begin{equation}
 p=-\frac{\mu  \sqrt{P_0}}{\sqrt{P_0}+1}\,.
 \end{equation}
This solution cannot represent a star as the metric is singular in the centre. In addition, the equation of state is not the one of a standard fluid which makes the solution not necessary desirable. However the above result reminds us that the Einstein equations in the static and spherically symmetric case can represent a number of other systems. Indeed, the solution above could represent a universe filled with dark energy is in perfect (and possibly unstable) equilibrium with  with a (naked) singularity. 

As a second, more relevant, example, let us set
\begin{equation}
f=\frac{1}{4}(1+ 4P)~,
\end{equation}
in this way  the function $g$ and Eq.~\rf{EqPKM} become 
\begin{equation}
P_{,\rho }=P- 4 P^2~,\qquad g=\frac{(3-4 P) P}{4 P+1}~,
\end{equation}
which implies
\begin{equation}
P=\frac{e^{\rho }}{P_0+4 e^{\rho}}~,
\end{equation}
with $P_0>0$. Solving for $Y$ one has
\begin{equation}
Y=\frac{2 e^{\rho }}{P_0+4 e^{\rho }}~,
\end{equation}
which corresponds to the metric
\begin{subequations}
\begin{align}
& d s^2 = -A d t^2+ B d r^2 +r^2\,\big(d\theta^2+\sin^2\!\theta d\phi^2\big),\\
& A =A_0\left(P_0+\frac{r^2}{r_0^2}\right)^2~, \label{TolmanIV} \\
& B =\frac{P_0 r_0^2+8 r^2}{P_0 r_0^2+4 r^2}~, 
\end{align}
\end{subequations}
together with the  energy density and pressure given by
\begin{subequations}
\begin{eqnarray}
&& \mu= \frac{4 \left(3 P_0 r_0^2+8
   r^2\right)}{\left(P_0 r_0^2+8 r^2\right){}^2}~,\\
&&p=\frac{4}{P_0 r_0^2+8 r^2}~,
\end{eqnarray}
\end{subequations}
 and the equation of state 
 \begin{equation}
\mu = 2 P_0r_0^2p^2+p~.
 \end{equation}
This solution is related to the so-called Tolman IV solution~\cite{Tolman:1939jz}. The difference is in the $(r,r)$ part of the metric. As we will see in the following this is not a mere coincidence.

The first and the third  solutions above give us two interesting solution prototypes. The first has a certain range of parameters for which  the central pressure is infinite. This is an indication that there exist a definite region in the parameter space in which these objects can exist and another in which they are unstable. The other, instead, has a Newtonian limit that is the same of the standard solution for the gravitational field in a Newtonian sphere of fluid of constant density. For this solution, no value of the parameters makes the central pressure to diverge and, therefore, although not necessarily stable, can be thought as ``more stable'' than the previous one.  We will make use of the two solutions above as template for the rest of our discussion. 
\section{Reconstructing physically relevant solutions}
Eqs.~\rf{112TOVPF}, although physically clear, are not very easy to be solved. We can better appreciate the reason behind this difficulty and generate some interesting solutions, using the reconstruction point of view proposed in general in Ref.~\cite{Carloni:2014rba}. Reconstruction algorithms have been used repeatedly in literature to generate new solutions (see {\it e.g.} Refs.~\cite{Lucchin:1984yf,Lidsey:1995np}). Here we propose an algorithm that allows  the derivation of two interesting generalisations of known solutions.

Starting from the Eqs.~(\ref{NewEqGenV1}-\ref{NewConstrGenV3}) we obtain
\begin{align}
&\mathbb{M}=\frac{\mathcal{K}_{,\rho }}{2 \mathcal{K}}+\mathcal{K}-\frac{1}{4}~,\\
& \nn P=
   \frac{1}{3} \left(2 Y_{,\rho }+2 Y^2+ Y\right)-\frac{2 Y+1}{6} \frac{
   \mathcal{K}_{,\rho }}{\mathcal{K}}\\
&~~~~~-\frac{1}{3} \mathcal{K}+\frac{1}{12}~,\\
&\nn 0=(2 Y+1) \mathcal{K}_{,\rho }-4 \mathcal{K}^2\\
&~~~~~-\mathcal{K} \left[4
   Y_{,\rho }+4 (Y-1) Y-1\right]~. \label{ISO_constr}
\end{align}
This form of the equations clearly shows the difficulty behind the resolution of the TOV equations in the homogeneous case. The constraint which relates the metric coefficients makes it difficult to find solutions for a given form of the energy density and/or the pressure. 

In reconstructing solutions for this case it is useful to keep control of the equation of state of the matter source. It is not too difficult to express it in terms of the variable $Y$ and $\K$ as:
\begin{equation}
\begin{split}
w=&\left[Y (2 Y+1) \left(4 \mathcal{K}+8 Y^2+4 Y_{\rho }-2 Y-1\right)\right]\times \\
& \left\{(-8 Y-4) Y_{,\rho \rho }+24 Y_{,\rho }^2 \right.\\
&+
4 \left(6 \mathcal{K}+6 Y^2-9 Y-1\right) Y_{\rho }\\
&\left.+Y [4 Y (6 \mathcal{K}+Y (4 Y-7)+1)+3-4 \mathcal{K}]\right\}^{-1}\,.
\end{split}
 \end{equation}
Once the form of the metric has been chosen it becomes straightforward from the expression above to determine if the source has an acceptable thermodynamics.

Let us now consider the coefficient $A$ of the solution in Eq.~\rf{TolmanIV}. Setting 
\begin{equation}
 A =A_0\left(a+b e^{\rho}\right)^2\,,
\end{equation}
one obtains
\begin{equation}
Y=\frac{b e^{\rho }}{a+b e^{\rho }}\,,
\end{equation}
and from Eq.~\rf{ISO_constr} it follows that
\begin{equation}
\K=\frac{\mathcal{K}_0\left(a+3 b e^{\rho }\right)^{2/3}}{e^{\rho }-4 \left(a+3 b e^{\rho
   }\right)^{2/3}}\,.
\end{equation}
Setting $\mathcal{K}_0=-1$  we arrive at the solution
\begin{subequations}\label{ReconISOMetric}
\begin{align}
& d s^2 = -Ad t^2+ B d r^2 +r^2\,\big(d\theta^2+\sin^2\!\theta d\phi^2\big),\\
& A =A_0\left(a+b\frac{r^2}{r_0^2}\right)^2,\\
& B =\left(1-\frac{ r^2}{4 r_0^{10/3} \left(a r_0^2+3 b r^2\right){}^{2/3}}\right)^{-1}.
\end{align}
\end{subequations}
The  energy density and the pressure are given by
\begin{equation}\label{ReconISOmup}
\begin{split}
&\mu=\frac{3 a r_0^2+5 b r^2}{4 r_0^{10/3} \left(a r_0^2+3 b r^2\right){}^{5/3}}~,\\
& p= \frac{16 b r_0^{10/3} \left(a r_0^2+3 b r^2\right){}^{2/3}-a r_0^2-5 b r^2}{4 r_0^{10/3} \left(a r_0^2+b r^2\right) \left(a r_0^2+3 b
   r^2\right){}^{2/3}}~,
\end{split}
\end{equation}
while the  barotropic factor is
\begin{align}\label{RecISOw}
 \nn\frac{\partial p}{\partial \mu}=w(r)=& \frac{\left(a+3 b r^2\right) \left(a^2-5 b^2 r^4\right)}{5 \left(a+b r^2\right)^3}\\
&\nn+\frac{8 a^2 b \left(a+3 b r^2\right)^{2/3}}{5
   \left(a+b r^2\right)^3}\\
&\nn +\frac{72 b^3  r^4 \left(a+3 b
   r^2\right)^{2/3}}{5\left(a+b r^2\right)^3}\\
& +\frac{48 a b^2  r^2
   \left(a+3 b r^2\right)^{2/3}}{5\left(a+b r^2\right)^3}~,
\end{align}
which is decreasing and takes only values between zero and one. 

\begin{figure}[!ht]
    \subfloat[The coefficients of the metric in Eq.~\rf{ReconISOMetric}. The blue line represents $A$ and the orange $B$. \label{F:ReconISOMetric}]{%
      \includegraphics[width=0.45\textwidth]{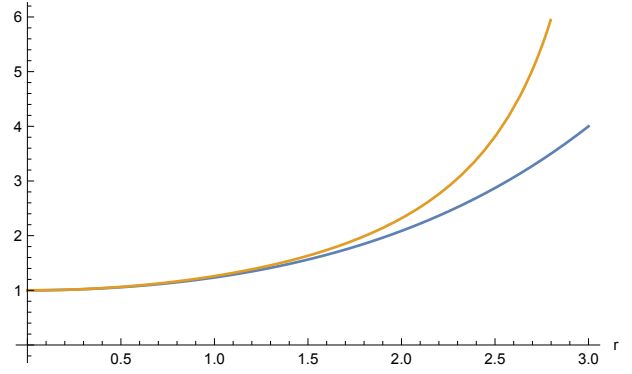}}\\\mbox{}\\
    \subfloat[The thermodynamic quantities in Eqs.~\rf{ReconISOmup}  associated with Eq.~\rf{ReconISOMetric}.The blue line represents  $p$ and the orange $\mu$.\label{F:ReconISOmup}]{%
      \includegraphics[width=0.45\textwidth]{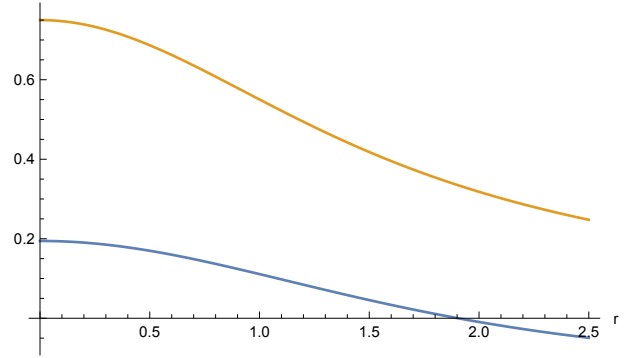}}
      \\\mbox{}\\
    \subfloat[The barotropic factor in Eq.~\rf{RecISOw} associated with Eqs.\rf{ReconISOmup}. Notice that the values of $w$ are compatible with a mix of a pressureless fluid and photons. \label{F:ReconISOw}]{%
      \includegraphics[width=0.45\textwidth]{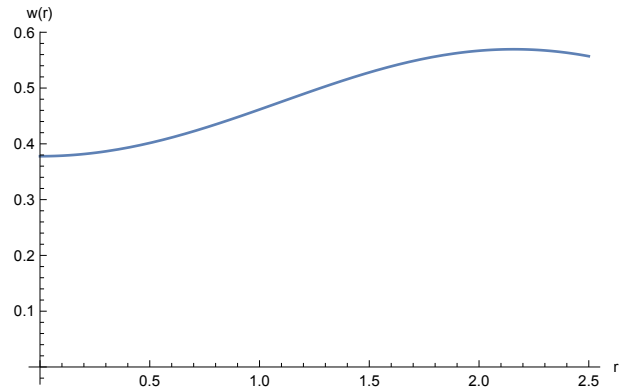}}
      \caption{Graphs of the key quantities of the solution in Eq.~\rf{ReconISOMetric} in the case $r_0=1$, $a=1/2$, $b=1/2$, $A_0=1$. }
          \label{F:RecISO}
  \end{figure}

In Fig.~\ref{F:RecISO} an example of the behaviour of the geometry and the thermodynamics of this solution is plotted for specific values of the parameters. With this choice of parameters the solution appears to present not only  decreasing energy density and pressure, but also a natural surface for the star ($p=0$).

One can generalise the reasoning above setting
\begin{equation} \label{RecAgen}
A=A_0\left(a+b\frac{r^2}{r_0^2}\right)^\beta\,,
\end{equation}
which for different values of $\beta$ reproduces  solutions like Durg IV, Durg V, Heint IIa, Heint IIIa, Heint IIIe in Ref.~\cite{Delgaty:1998uy}.  As we will see in the next Section the reason behind this similarities is related to the existence of some theorems which connect different solutions of the TOV equations.

Let us now  try to reconstruct a solution in which the metric coefficient $A$ is given by Eq.~\rf{MuConstFlA}.

\begin{figure}[!ht]
    \subfloat[The coefficients of the metric in Eq.~\rf{ReconISOMetric2}. The blue line represents $A$ and the orange $B$. \label{F:ReconISOMetric2}]{%
      \includegraphics[width=0.45\textwidth]{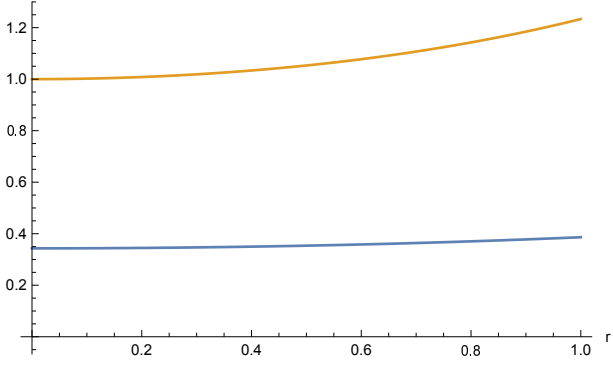}}\\\mbox{}\\
    \subfloat[The thermodynamic quantities in Eqs.~\rf{ReconISOmu2} and ~\rf{ReconISOp2} associated with Eq.~\rf{ReconISOMetric2}  in a semilogarithmic plot
    . The blue line represents  $p$ and the orange $\mu$.\label{F:ReconISOmup2}]{%
      \includegraphics[width=0.45\textwidth]{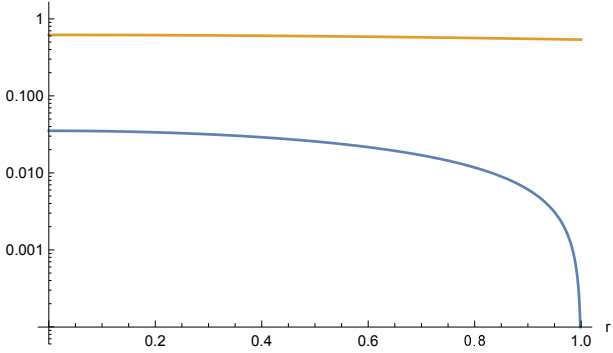}}
      \\\mbox{}\\
    \subfloat[The barotropic factor in Eq.~\rf{ReconISOMetric2} associated with Eqs.~\rf{ReconISOmu2} and ~\rf{ReconISOp2}. Notice that the values of $w$ are compatible with a mix of a pressureless fluid and photons. \label{F:ReconISOw2}]{%
      \includegraphics[width=0.45\textwidth]{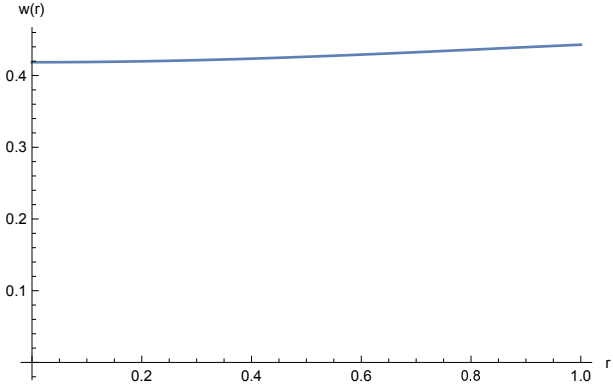}}
      \caption{Graphs of the key quantities of the solution in Eq.~\rf{ReconISOMetric2} in the case $r_0=1$,  $a=-2$, $b=1/10$, $c=2$, $A_0=1$. $\K_0$ has been chosen so that $p=0$ at $r=1$.}
          \label{F:RecISO2}
  \end{figure}

Setting 
\begin{equation}
 A =A_0 \left(a+\sqrt{b+c e^{\rho} }\right)^2~,
\end{equation}
which corresponds to
\begin{equation}
Y =-\frac{b e^{\rho }}{2 \sqrt{c-b e^{\rho }} \left(a+\sqrt{c-b e^{\rho }}\right)}~,
\end{equation}
 Eq.~\rf{ISO_constr} gives
\begin{equation}
\begin{split}
&\K =\frac{c \psi  \left(a \sqrt{c-b e^{\rho }}-2 b e^{\rho }+c\right)}{\left(c-b
   e^{\rho }\right) \left[4 \psi \left(a  \sqrt{c-b e^{\rho }}-2 b e^{\rho }+ c\right)-b c \mathcal{K}_0 e^{\rho }\right]}~,\\
&\psi =\left(\frac{\sqrt{a^2+8 c}+a+4 \sqrt{c-b e^{\rho }}}{\sqrt{a^2+8 c}-a-4 \sqrt{c-b e^{\rho }}}\right)^{-\frac{a}{\sqrt{a^2+8 c}}}~,  
\end{split}
\end{equation}
which leads, using the area radious $r$, to 
\begin{subequations}\label{ReconISOMetric2}
\begin{align}
& d s^2 = -Ad t^2+ B d r^2 +r^2\,\big(d\theta^2+\sin^2\!\theta d\phi^2\big),\\
& A =\left(a+\sqrt{c-\frac{b r^2}{r_0^2}}\right)^2,\\
& B =\-\frac{4r_0^2 c \left[r_0 \left(a \sqrt{c r_0^2-b r^2}+c r_0\right)-2 b r^2\right]}{
   \left(c r_0^2-b r^2\right)}\times\\
   &~~~~~~~~ \left[b r^2 \left(c \mathcal{K}_0 \psi +8\right)-4 r_0
   \left(a \sqrt{c r_0^2-b r^2}+c r_0\right)\right]^{-1
  }~,\\
& \psi =\left(\frac{\sqrt{a^2+8 c}+a+4 \sqrt{c r^{2 }_0-b r^{2 }}}{\sqrt{a^2+8 c}-a-4 \sqrt{c r^{2 }_0-b r^{2 }}}\right)^{-\frac{a}{\sqrt{a^2+8 c}}}~.
\end{align}
\end{subequations}
It is quite straightforward to realise that the new solution reduces to the one of Eq.~\rf{MuConstFl} for $\K_0=0$. Notice also that since the coefficient $B$ goes to a constant for $r=0$, one can set the constants in such a way to have $B(0)=1$ avoiding any conical singularity. However, differently from the case of Eq.~\rf{MuConstFl}, this solution does not have a constant energy density. Indeed one obtains 
\begin{equation}\label{ReconISOmu2}
\begin{split}
&\mu=\frac{b}{4 c \left(a+\sqrt{c-b r^2}\right) \left(a \sqrt{c-b r^2}-2 b r^2+c\right)^3}\times\\
&\Big\{12 a^3 \left(c-b r^2\right)^2\\
&+3 a^2 \left(c-b r^2\right)^{3/2} \left[c \mathcal{K}_0 \psi  \left(c-b r^2\right)+12 \left(c-2 b r^2\right)\right]\\
&+2 a
   \left(c-b r^2\right) \left[c \mathcal{K}_0 \psi  \left(6 b^2 r^4-8 b c r^2+3 c^2\right)\right.\\
&\left.+18 \left(c-2 b r^2\right)^2\right]\\
&+\left(c-2 b r^2\right)
   \sqrt{c-b r^2} \left[c \mathcal{K}_0 \psi  \left(6 b^2 r^4-7 b c r^2+3 c^2\right)\right.\\
&\left.+12 \left(c-2 b r^2\right)^2\right]\Big\}~,
\end{split}
\end{equation}
\begin{equation}\label{ReconISOp2}
\begin{split}
& p=\frac{b}{4 c \left(a+\sqrt{c-b r^2}\right) \left(a \sqrt{c-b r^2}-2 b r^2+c\right)^3}\times\\
&\Big\{-4 a^4 \left(c-b r^2\right)^{3/2}\\
&-a^3 \left(c-b r^2\right) \left[c \mathcal{K}_0 \psi  \left(c-b r^2\right)-36 b r^2+24 c\right]\\
&-a^2 \sqrt{c-b r^2}
   \left[c \mathcal{K}_0 \psi  \left(3 c-7 b r^2\right) \left(c-b r^2\right)\right.
  \\
&\left.+12 \left(4 c-5 b r^2\right) \left(c-2 b r^2\right)\right]\\
&-a \left(c-2 b
   r^2\right) \left[c \mathcal{K}_0 \psi  \left(3 c-8 b r^2\right) \left(c-b r^2\right)\right.\\
&\left.+4 \left(10 c-11 b r^2\right) \left(c-2 b
   r^2\right)\right)]\\
&+\left(c-2 b r^2\right)^2 \sqrt{c-b r^2} \left[24 b r^2 \right.\\
&\left.-c \left(\mathcal{K}_0 \psi  \left(c-3 b r^2\right)+12\right)\right]\Big\}~.
\end{split}
\end{equation}
With these expressions one can easily calculate the (rather lengthy)  barotropic factor of the fluid. In Fig.~\ref{F:RecISO2}  we give an example of the behaviour of the metric and the thermodynamical quantities. Notice that, with the parameter chosen, this solution posses a barotropic factor which corresponds to a mix of pressureless matter and radiation.

\section{Generating Theorems}
The  general properties of the solutions of the TOV equations have been studied in detail from different persectives. Recently in Refs.~\cite{Boonserm:2005ni,Boonserm:2006up} some theorems were proven which are able to map different exact solutions of the TOV equations into each other. 

We will show now that in our variables similar theorems can be defined considering a linear deformation of the quantities characterising a given solution. 

Let us consider for example, the case in which given a solution of Eq.~\rf{112TOVPF} indicated by $P_0,\mathbb{M}_0,\K_0, Y_0$ we perform the deformation 
\begin{equation}\label{TeoGen1}
P= P_0 + P_1\,,\qquad Y= Y_0 + Y_1\,.
\end{equation}
In terms of the metric coefficients, the transformation in Eq.~\rf{TeoGen1} leads to
\begin{subequations}\label{TeoGen1Coord}
\begin{align}
& A \rightarrow A_0 (\rho) \exp\left(\int Y_1 d\rho\right)~,\\
& B \rightarrow B_0 (\rho)~,\\
& C\rightarrow C_0 (\rho)~,
\end{align}
\end{subequations}
where now $A_0 (\rho), B_0 (\rho), C_0 (\rho)$ are the metric coefficients associated to the solutions $P_0,\mathbb{M}_0,\K_0, Y_0$ of Eq.~\rf{112TOVPF}. Notice that this transformation corresponds to theorem 2 of Ref.~\cite{Boonserm:2005ni}. 

Substituting in Eq.~\rf{112TOVPF} and using the constraint in Eq.~\rf{NewConstrGenV2} we obtain that $P_1$ and  $Y_1$ have to satisfy the following relations:
\begin{equation}\label{TeoGen1Eq}
\begin{split}
&P_{1, \rho}+P_1^2+P_1 \left(3 \mathcal{K}_0-\mathbb{M}_0+2 P_0-\frac{7}{4}\right)=0~, \\
& Y_1=P_1~.
\end{split}
\end{equation}
The first equation above is a Bernoulli equation whose formal general solution is:
\begin{equation} 
\begin{split}
&P_1=\frac{e^{F}}{ P_{*} + \int e^{F} d\rho}~, \\
&F= \int \left(3 \mathcal{K}_0-\mathbb{M}_0+2 P_0-\frac{7}{4}\right) d\rho~.
\end{split}
\end{equation}
Setting $P_1=u_{,\rho}/u$, Eq. \eqref{TeoGen1Eq} reduces to : 
\begin{equation}\label{TeoGen2Eq}
u_{,\rho\rho}+ u_{,\rho} \left(3 \mathcal{K}_0-\mathbb{M}_0+2 P_0-\frac{7}{4}\right)=0,
\end{equation}
which, since $Y_1=P_1$ and given the expression of $Y$ in terms of the metric coefficient $A$, matches exactly the key equation obtained in theorem 2 of Ref.~\cite{Boonserm:2005ni}. Hence, solving either of \eqref{TeoGen1Eq} and \eqref{TeoGen2Eq}, one finds a new solution which corresponds to the transformed metric coefficients of \eqref{TeoGen1Coord}.

In the same way one can obtain theorems corresponding to the deformations of other parameters (although the calculations are slightly more complicated). In the case of the deformation
\begin{equation}\label{TeoGen2}
\begin{split}
&P= P_0 + P_1,\qquad \M= \M_0 + \M_1~,\\
&\K= \frac{1}{\K_0+\K_1}~,
\end{split}
\end{equation}
which leads to 
 \begin{subequations}
\begin{align}
& A \rightarrow A_0 (\rho)~,\\
& B^{-1} \rightarrow B_0 (\rho) + \frac{e^\rho}{K_0}\K_1~,\\
& C\rightarrow C_0 (\rho)~,
\end{align}
\end{subequations}
and corresponds to the theorem 1 in Ref.~\cite{Boonserm:2005ni} (see also~\cite{ExactSolutions}). 

Eq.~\rf{112TOVPF}, the constraint in Eq.~\rf{NewConstrGenV2} and Eq.~\rf{NewEqGenV1} return 
\begin{equation}
\begin{split}
&\mathcal{K}_{1,\rho}=-\mathcal{K}_1 \Phi+\Gamma~,\\
& \Phi=\frac{12 \mathcal{K}_0-4 \mathbb{M}_0-1+Y_0 \left(8 \mathcal{K}_0-8 \mathbb{M}_0-2\right)}{ 2(1+2 Y_0)}~,\\
& \Gamma= \frac{8 \mathcal{K}_0^2\left(Y_0+1\right)+ \mathcal{K}_0 \left(4 \mathbb{M}_0+1\right) \left(2
   Y_0+1\right)-4}{ 1+2 Y_0}~,\\
& P_1=\frac{\mathcal{K}_0^2+\mathcal{K}_1 \mathcal{K}_0-1}{\mathcal{K}_0+\mathcal{K}_1}~,\\
&\M_1=-\frac{\left[\mathcal{K}_0 \left(\mathcal{K}_0+\mathcal{K}_1\right)-1\right] \left(2
   Y_0+3\right)}{\left(\mathcal{K}_0+\mathcal{K}_1\right) \left(2 Y_0+1\right)}~.
\end{split}
\end{equation}
As before, the first equation above is a linear differential equation which can always be solved formally as:
\begin{equation}
\begin{split}
&\K_1=e^{-F} \left(\K_{*} -\int e^{F} \Gamma d\rho\right)~, \\
&F= \int \Phi d\rho~,
\end{split}
\end{equation}
which proves the theorem.

The $1+1+2$ equations reveal the presence of a number of additional deformation theorems (as well as no go theorems) for non linear deformations of the solutions.  For example, the last theorem above can be further generalised. Setting
\begin{equation}\label{TeoGen2}
\begin{split}
&P= P_0 + P_1,\qquad \M= \M_0 + \M_1~,\\
&\K= \K_1~,
\end{split}
\end{equation}
where $\K_1$ is a generic function of $\rho$ (and therefore even a function of $\K_0$). This deformation corresponds to 
 \begin{subequations}
\begin{align}
& A \rightarrow A_0 (\rho)~,\\
& B \rightarrow \frac{ e^\rho}{K_0}\K_1~,\\
& C\rightarrow C_0 (\rho)~.
\end{align}
\end{subequations}
Under \eqref{TeoGen2},Eq.~\rf{112TOVPF}, the constraint in Eq.~\rf{NewConstrGenV2} and Eq.~\rf{NewEqGenV1} return:
\begin{equation}
\begin{split}
& \mathcal{K}_{1,\rho}=\frac{4 \mathcal{K}_1^2}{2 Y_0+1}+\mathcal{K}_1 \left(2 \mathbb{M}_0-\frac{2
   \mathcal{K}_0 \left(2 Y_0+3\right)}{2 Y_0+1}+\frac{1}{2}\right)~, \\
& P_1=\mathcal{K}_0-\mathcal{K}_1~,\\
&\M_1=-\frac{\left(\mathcal{K}_0-\mathcal{K}_1\right) \left(4 \mathcal{K}_0+4 P_0+5\right)}{4\mathcal{K}_0+4 P_0+1}~.
\end{split}
\end{equation}
The first equation above is a Bernoulli differential equation which can always be solved exactly as:
\begin{equation}
\begin{split}
&\K_1=\frac{e^{F}}{ \K_{*} + \int e^{F} G d\rho}~, \\
&F= \int \frac{4 d\rho}{2 Y_0+1}~, \\
& G=\frac{2
   \mathcal{K}_0 \left(2 Y_0+3\right)}{2 Y_0+1}-2 \mathbb{M}_0-\frac{1}{2}~.
\end{split}
\end{equation}

These theorems can be easily understood in terms of the constraints in Eqs.~\eqref{NewConstrGenV1}-\eqref{NewConstrGenV2}. For example, it is clear from Eq.~\rf{NewConstrGenV2} that a deformation of the pressure variable cannot leave $Y$ and $\K$ (i.e. the metric coefficients) both unchanged. And yet, different metrics for which the quantity $Y-\K$ remains unchanged can correspond to a single pressure profile. Finally, as remarked in Refs.~\cite{Boonserm:2005ni,Boonserm:2006up}, these different theorems can be combined to obtain chains of exact solutions.

\section{Conclusions}
In this paper we have used the $1+1+2$ covariant formalism and the variables in Ref.~\cite{Carloni:2014rba} to describe non vacuum static spherically symmetric spacetimes and give a covariant generalisation of the TOV equations in the case of an isotropic fluid.

The new formalism offers a different point of view on the structure of relativistic stellar objects that simplifies a number of aspects of the TOV equations, some of which have been recently pointed out in literature. It also gives a very simple description of Israel's junction conditions clarifying the relation between the requirements on the continuity of the metric and the behaviour of the thermodynamical quantities characterising the fluids. 

The covariant formulation of the TOV equations shows that when they are solved assigning a density profile, one needs to solve a system of a Bernoulli and a Riccati equation to obtain the desired solutions. This fact implies that except for some special cases~\cite{Mak:2013pga}, we cannot write a general analytic solution of the TOV equations in terms of elementary functions, although we can look at the properties of such solutions (see {\it e.g.} Ref.~\cite{Baumgarte:1993}). However, the difficulty of the mathematical problem also depends on the resolution strategy. For example, assigning a non-trivial equation of state might imply the resolution of  even more complex differential equations (e.g Abel equations).

The covariant TOV equations that we have presented are also useful for the development of a complete reconstruction algorithm. The structure of such algorithm clearly shows the difficulty of obtaining exact solutions in for isotropic stellar objects: the metric coefficients appear to be related by a differential constraint. The algorithm also allows to connect the metric coefficients (and their derivatives) to the barotropic factor of the fluid. This result helps the control of the choice of the solution to reconstruct.
We presented two different solutions obtained by using this algorithm. The first is characterised by a $(0,0)$ component of the metric that resembles the Tolman IV solution. The second, instead, is constructed in such a way to present the same instability of the classical constant density solution, but with a non-constant $\mu$. Both the solutions we found present at least a combination of parameters for which they satisfy all the physical conditions given in Section \ref{112TOV} (see Figs.~\ref{F:RecISO} and~\ref{F:RecISO2}). It is interesting to notice that the second solution is characterized, in the parameter range that we have explored, by a density which is almost constant and this feature makes it a relevant generalisation of the known solutions of this type. 

Finally, we used the covariant TOV equations to rederive the generating theorems of Refs.~\cite{Boonserm:2005ni,Boonserm:2006up,ExactSolutions}. By means of the new variables these theorems can be obtained as a linear deformations of a given solution. It is also clear that the constraints in Eqs.~(\ref{NewConstrGenV1})-(\ref{NewConstrGenV2}) can be used as a guide to find the possible theorems that can be found. For example, Eq.~\rf{NewConstrGenV2} shows that a shift in $Y$ (and therefore in $A$) can be consistent with a shift in $\K$ (and therefore in $B$)  with $p$ and $\mu$ fixed, or a shift in $p$ and $\K$ and $\mu$ fixed.

The presence of the generating theorems gives us a deeper understanding of the results that we have obtained via the reconstruction technique. For example, Eq.~\rf{ReconISOMetric} can be obtained by Eq.~\rf{TolmanIV} and the other solutions generated via Eq.~\rf{RecAgen} are connected to some of the solutions in Ref.~\cite{Delgaty:1998uy}.

To conclude, the covariant version of the TOV equations constitute a powerful tool for the investigation of the structure of stellar objects. The advantages of this framework will become even more clear when applied to  more complicated cases, like the one in which the source is an anisotropic fluid. A future work will deal with such case in detail.

\begin{acknowledgments}
SC and DV were supported by the Funda\c{c}\~{a}o para a Ci\^{e}ncia e Tecnologia through project IF/00250/2013 and acknowledge financial support provided under the European Union's H2020 ERC Consolidator Grant ``Matter and strong-field gravity: New frontiers in Einstein's theory'' grant agreement No. MaGRaTh646597.
\end{acknowledgments}

\end{document}